\documentclass[draft]{agujournal2019}
\usepackage{url} 
\usepackage{lineno}
\usepackage[finalnew]{trackchanges} 
\usepackage{soul}
\draftfalse

\journalname{JGR: Planets}

\begin{document}

%
%

\title{Compositional Measurements of Saturn's Upper Atmosphere and Rings from Cassini INMS}

%
%



\authors{J. Serigano\affil{1}, S. M. H{\"o}rst\affil{1}, C. He\affil{1}, T. Gautier\affil{2}, R. V. Yelle\affil{3}, T. T. Koskinen\affil{3}, \add{M. G. Trainer}\affil{4}}
\affiliation{1}{Department of Earth and Planetary Sciences, Johns Hopkins University, Baltimore, MD, USA.}
\affiliation{2}{LATMOS-IPSL, CNRS, Sorbonne Universit{\'e}, UVSQ, Guyancourt, France}
\affiliation{3}{Lunar and Planetary Laboratory, University of Arizona, Tucson, AZ, USA.}
\affiliation{4}{NASA Goddard Spaceflight Center, Greenbelt, MD, 20771, USA.}



\correspondingauthor{Joseph Serigano}{jserigano4@jhu.edu}




\begin{keypoints}
\item We measure the density profiles of H$_2$, He, CH$_4$, H$_2$O,  and NH$_3$ in Saturn's \change{upper atmosphere}{thermosphere} from Cassini INMS. 
\item We use a new mass spectral deconvolution algorithm to determine the relative abundances of different species found in the mass spectra. 
\item We report further evidence of CH$_4$, H$_2$O, and NH$_3$ entering Saturn's atmosphere from the rings at a rate of at least 10$^3$ kg/s.
\end{keypoints}

%
%

%
%

\begin{abstract}
The Cassini spacecraft's last orbits directly sampled Saturn's \change{upper atmosphere}{thermosphere} and revealed a much more chemically complex environment than previously believed. Observations from the Ion and Neutral Mass Spectrometer (INMS) aboard Cassini provided compositional measurements of this region and found an influx of material raining into Saturn's upper atmosphere from the rings. We present here an in-depth analysis of the CH$_4$, H$_2$O, and NH$_3$ signal from INMS and provide further evidence of external material entering Saturn's atmosphere from the rings. We use a new mass spectral deconvolution algorithm to determine the amount of each species observed in the spectrum and use these values to determine the influx and mass deposition rate for these species. 
\end{abstract}

\section*{Plain Language Summary}
The Cassini spacecraft's last orbits around Saturn provided measurements to help us understand how the rings of Saturn interact with its upper atmosphere. Using measurements from the mass spectrometer aboard the spacecraft, we find that a lot of material from the rings is entering Saturn's atmosphere. We use a new method to determine the amount of water, methane, and ammonia that are entering the atmosphere from the rings and find that this large influx could be deplete the ring system in a relatively short amount of time.

%
%

%


%
%
%
%

\section{Introduction}
In September 2017, the Cassini-Huygens mission to the Saturn system came to an end as the spacecraft intentionally entered the planet's atmosphere. Prior to entry, the Cassini spacecraft spent its last five months executing a series of 22 highly inclined "Grand Finale" orbits through the previously unexplored region between Saturn and its extensive ring system, yielding the first ever direct sampling of this region and the planet's upper atmosphere. The unique trajectory of these orbits along with the spacecraft's proximity to Saturn allowed for unprecedented studies of the planet's complex interactions with the surrounding environment. During these orbits, Cassini obtained measurements near the equatorial ring plane at various heights above the planet's 1-bar pressure level. The final five of these orbits reached the lowest altitudes and directly sampled Saturn's upper thermosphere while orbits prior to these sampled the region between the planet and its innermost D ring. The spacecraft's last encounter, known as the "Final Plunge," represents the deepest sampling of Saturn's atmosphere and provided measurements down to approximately 1370 km above the 1-bar pressure level before losing contact with Earth. 

The Grand Finale data returned from the spacecraft's last months have already revolutionized our understanding of this unique region in our solar system, highlighting ring-planet interactions like never before. \citeA{mitchell18} used measurements from Cassini's Magnetospheric Imaging Instrument (MIMI) to conclude that interactions between the upper atmosphere and inner edge of the D ring resulted in small dust grains entering Saturn's atmosphere from the rings in a narrow region near the equatorial ring plane. Using the Cosmic Dust Analyzer (CDA), which is sensitive to larger particles than MIMI,  \citeA{hsu18} observed a greater influx of exogenous material with a greater latitudinal spread \add{than the MIMI results} that reached into Saturn's mid latitude region. Ionospheric fluctuations as a consequence of ring-atmosphere coupling have also been observed. Using electron density measurements from the Radio and Plasma Wave Science (RPWS) instrument, \citeA{wahlund18} reported a highly variable ionosphere with large decreases in ionization in regions where the planet is shadowed by the rings. Additionally, \citeA{cravens18} used measurements from the ion mode of the Ion and Neutral Mass Spectrometer (INMS) to conclude that the composition of light ion species in Saturn's ionosphere can only be explained by the presence of heavier molecular species coming from the rings.

Although the measurements from Cassini's final orbits redefined our understanding of ring-atmosphere interactions, the idea of "ring rain", where external material from the rings enters Saturn's upper atmosphere, has existed for decades and evidence for this phenomenon was first reported during the Voyager era. Using radio occultation data from the Pioneer and Voyager spacecraft, \citeA{connerney84} proposed the first ionospheric model of Saturn in which exogenous charged water particles from the rings were used to explain the observed electron density at Saturn, which was an order of magnitude less than models predicted. More recently, using ground based observations from the Keck telescope, \citeA{odon13} and \citeA{odon17} found significant variations in Saturn's midlatitude H$_3$$^+$ intensity that could not be explained by solar activity. Instead, they concluded that these variations can be attributed to the transport of charged species derived from water via regions of the rings that are magnetically linked to the atmosphere.

Cassini's Ion and Neutral Mass Spectrometer (INMS) provided the first in situ compositional measurements of Saturn's upper atmosphere during the Grand Finale. Two of the instrument's main objectives at this time were to determine the abundances of H$_2$ and He, the major constituents of Saturn's atmosphere, and to characterize any interaction between the upper atmosphere and the rings. These measurements were taken well above Saturn's homopause, the level below which an atmosphere is well-mixed and assumes a scale height  in accordance with the mean mass of an atmospheric molecule. Above the homopause, molecular diffusion is a mass dependent process. Thus, a molecule whose mass is heavier than the bulk atmosphere will decrease in abundance above the homopause more rapidly than a lighter molecule. As a result of this mass-dependent diffusive separation, models of Saturn's upper atmosphere suggested that only the abundant lighter species H$_2$ and He would be detected, with other heavier minor species being well below the instrument's detection limit (see e.g., \citeA{koskinen15}). However, the compositional measurements from INMS revealed a surprisingly large amount of heavier constituents influencing the upper atmosphere, with the evidence again suggesting that a material influx from the rings is the most likely source of these heavier molecules. 

\citeA{yelle18} used measurements from the neutral mode of INMS to determine the density profiles of low mass constituents and the neutral temperature profile in Saturn's upper atmosphere. They reported the distributions of H$_2$, He, and CH$_4$ in Saturn's equatorial thermosphere, noting that the density profiles of H$_2$ and He are consistent with an atmosphere in diffusive equilibrium, but that the higher than expected amounts of CH$_4$ in this region of the atmosphere can only be attributed to an external source with a flux of approximately 1.2 $\times$ 10$^{13}$ m$^{-2}$s$^{-1}$. \citeA{waite18} and \citeA{miller20} analyzed a larger portion of the instrument's mass range, confirming the existence of heavy neutral hydrocarbons infalling from the rings into the atmosphere and attributing the influx to certain regions of the D ring. \citeA{waite18} and \citeA{perry18} both concluded that at least 10$^4$ kg/s of endogenous material was being deposited into Saturn's upper atmosphere from the rings during these observations. This surprisingly large and unsustainable influx of material led the authors to conclude that this influx must be transient and a result of a recent perturbation in the D ring causing an unusually high amount of material to fall into the planet. 

In this study, we expand on the neutral INMS results first presented in \citeA{yelle18} to include other species with external origins. We use a new sophisticated mass spectral deconvolution algorithm for interpreting mass spectra returned by spacecraft as detailed in \citeA{gautier19}. The species of interest for this study are H$_2$O, CH$_4$, and NH$_3$. We limit our scope to these three species due to their importance in the outer solar system and in understanding ring composition and in order to focus on a segment of the mass spectrum where the signature of these three molecules overlap significantly (\textit{m/z} 12 to 20 amu).

\section{Instrument and Observations}

Measurements presented in this paper rely on data from the Closed Source Neutral (CSN) mode of the Ion and Neutral Mass Spectrometer (INMS) aboard the Cassini spacecraft. The primary focus of INMS was to characterize the composition, density, and temperature structure of Titan's upper atmosphere and its interaction with Saturn's magnetospheric plasma. A detailed description of the instrument can be found in \citeA{waite04}. The instrument's excellent performance throughout the spacecraft's 13 years in orbit allowed for a large number of studies that drastically improved our understanding of Titan's atmosphere. The instrument also directly sampled the plumes of Enceladus multiple times beginning in 2008. Thus, a detailed understanding of the instrument and how it functions in various environments found in the Saturn system already exists (see e.g., \citeA{waite05, waite07, waite09, cui08, cui09a, cui09b, teolis10, cui12, waite17}). 

The CSN mode of this instrument measures neutral species by first ionizing the molecules. This fragments each molecule into a characteristic pattern and the unique spectral signature of the resulting ionized fragments are then detected by the instrument in order to determine the composition of the inflowing sample. The inflowing material enters a spherical antechamber and travels to an ionization region where it is ionized by a collimated electron beam at 70 eV. The resulting ions are deflected onto the instrument's detectors by a dual radio frequency quadrupole mass analyzer, which filters the ions according to their mass-to-charge (\textit{m/z}) ratio. The instrument's dual detector system is electronically biased, with the majority of ions deflected onto the primary detector and a small fraction making it to the low gain secondary detector, which is utilized only in instances when the count rate of the primary detector saturates. Data are recorded in mass channels from 1 to 99 atomic mass units (amu) with a resolving power of 1 amu.  

This paper focuses on measurements taken during Cassini orbits 288 to 293, the lowest altitude passes of Cassini's Grand Finale orbits which directly sampled Saturn's thermosphere. These data can be found in the Planetary Plasma Interactions (PPI) node of the NASA Planetary Data System (PDS) public archive (https://pds-ppi.igpp.ucla.edu) \cite{waite_inms}. Orbits 288 to 292 sampled Saturn's atmosphere down to an altitude of about 1600 to 1700 km above Saturn's 1-bar pressure level. Orbit 289 was not optimized for INMS observations and will not be discussed in our analysis. Orbit 293, which included atmospheric entry, provided measurements down to approximately 1370 km above the 1-bar pressure level. INMS measurements in mass channel 2 are used to determine the H$_2$ density in the atmosphere and were taken every $\sim$0.6 s around closest approach. Measurements in other mass channels of particular interest were taken every $\sim$1 s. The spacecraft's velocity during these orbits was approximately 30 km/s. This corresponds to a spatial resolution of 18 km and 30 km, respectively, along the spacecraft trajectory. Additional orbital information can be found in Table \ref{tab:orbitalinfo}.

\begin{table}
\caption{Orbital information for the measurements used in this study. These were the last orbits of the Cassini spacecraft and represented the lowest altitude sampling of Saturn's upper atmosphere. Cassini entered Saturn's atmosphere during orbit 293.}
\centering
\begin{tabular}{c c c c c c c c}
\hline
Orbit & Date & Latitude & Longitude & LST & Altitude & Grav. Potential \\
Number & & ($^{\circ}$) & ($^{\circ}$) & (hr) & (km) & ($\times$ 10$^8$ J kg$^{-1}$) \\
\hline
288 & 14 August 2017 & -4.8 & -167.5 & 11.7 & 1705 & 6.695\\
290 & 27 August 2017 & -4.9 & 160.5 & 11.5 & 1626 & 6.702 \\
 291 & 2 September 2017 & -5.0 & -34.0 & 11.4 & 1639 & 6.700 \\
 292 & 9 September 2017 & -5.0 & 130.4 & 11.4 & 1675 & 6.697\\
 293 & 15 September 2017 & 9.2 & -54.3 & 10.8 & 1367 & 6.722 \\
\hline 
\end{tabular}\\
\footnotesize{\textit{Note.} Values in this table correspond to the spacecraft's closest approach to Saturn. Spacecraft velocity at this time was between 30.1 to 31 km/s for these orbits. LST = local solar time.}\\
\label{tab:orbitalinfo}
\end{table}

\section{Methods}
\subsection{Data Reduction}

Although INMS has been extensively characterized for studies of Titan's N$_2$-dominated atmosphere (see e.g., \citeA{yelle06, yelle08, muller08, cui08, cui09a, cui09b, magee09, cui12}), it is crucial to ensure that the instrument is still behaving as expected in Saturn's H$_2$-dominated atmosphere. Thus, data reduction is an especially important procedure for this particular data set since the instrument was performing in a new environment for which it was not designed. Additionally, the spacecraft orbited Saturn at speeds approximately 5 times faster than typical Titan flybys during these final orbits. This could impart excess energy into the system which could dissociate molecules entering the instrument's antechamber before ionization. Thus, it is likely that a fraction of the signal recorded by the instrument is from fragments of larger particles outside of the instrument's mass range. However, as reported in \citeA{waite18}, the abundance of larger organics seen in the spectra is at least an order of magnitude lower than that of CH$_4$ and would not have a very significant impact on the CH$_4$ abundance. They also report that the H$_2$ profile derived from INMS before and after closest approach suggests that effects from the spacecraft's speed are negligible in close proximity to the planet. Other recent studies also find no significant effects on the measurements from the spacecraft's high speed \cite{yelle18, perry18, miller20} and our analysis thus far shows no evidence of issues stemming from the spacecraft's speed. 

Many factors affect the response of the instrument. These factors have been extensively characterized for Titan's atmosphere and methods for correction are detailed in \citeA{cui08},  \citeA{magee09}, \citeA{cui09a}, and \citeA{cui12}. We adopt similar methods here and recharacterize the corrections to ensure that they are suitable for Saturn conditions. These include background subtraction and corrections for detector dead-time, calibration sensitivity, ram pressure enhancement, and counter saturation, which are briefly explained below.

\add{Residual gas present in the INMS chamber is responsible for outgassing and enhancing the signal in certain mass channels. Far from Saturn, this background tends to a constant level which must be properly subtracted in order to remove this enhancement. To determine the mean background signal for each orbit we average data taken well before closest approach to Saturn where signal in mass channel 2 has not yet begun to increase, which would indicate the detection of Saturn's extended atmosphere. The radiation background, which is a mass independent enhancement in signal that is due mainly to detection of charged particles from Saturn's magnetosphere, must also be removed. We determine the radiation background using the signal in mass channels 5 to 8, where no signal is expected and thus the only signal detected here is due to external radiation. The region of the mass spectrum analyzed in this study has a very strong signal in Saturn's atmosphere, thus the background effects are not very significant. However, proper removal of the background signal is critical and is even more crucial when analyzing the signal in higher mass channels where signal is much lower but still present.}

\begin{figure}
\noindent\includegraphics[width=\textwidth]{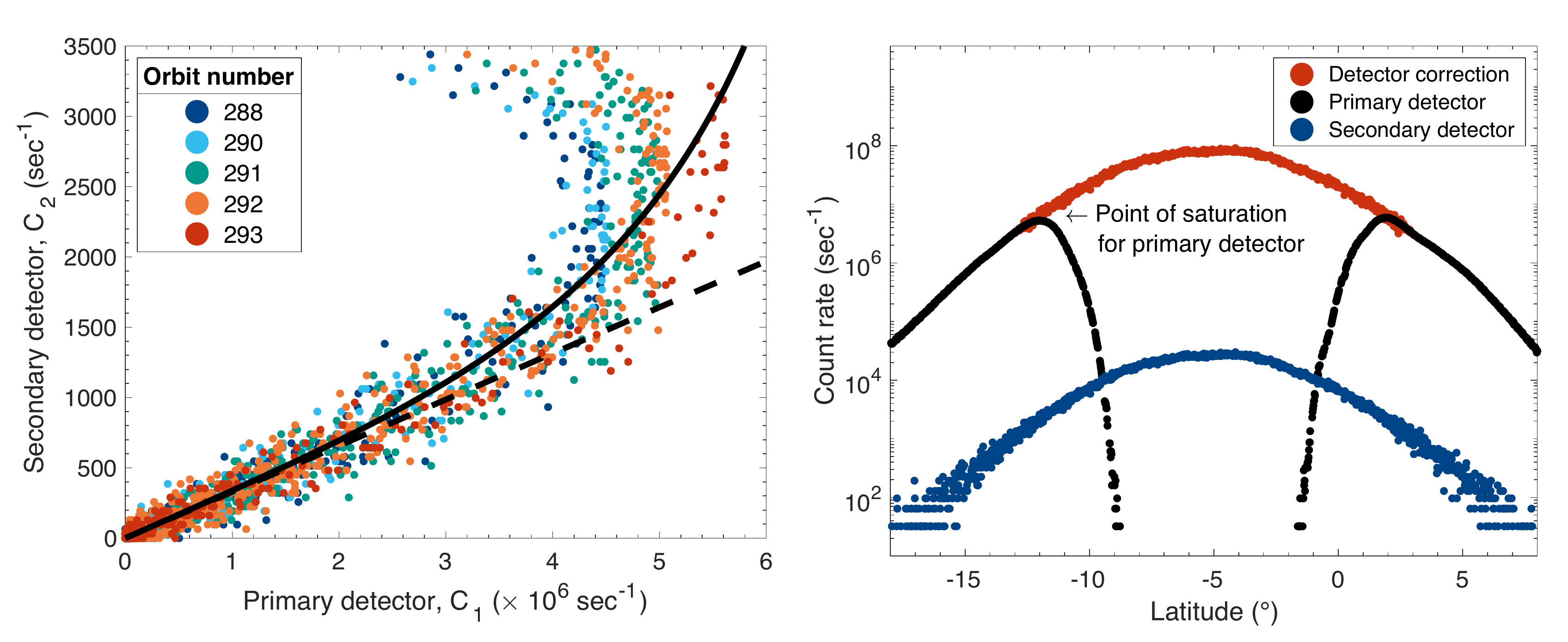}
\caption{Left: Count rate in mass channel 2 (H$_2$) from the secondary detector (C$_2$) as a function of count rate in mass channel 2 from the primary detector (C$_1$) for all five orbits analyzed here. The dashed line represents the linear correlation between these detectors at lower count rates and is the trend that the signal would follow if the primary detector was not affected by saturation. The solid line represents the nonlinear empirical relationship (equation 1) used to correct for saturation in the primary detector. Doing so increases the signal to noise ratio in our results and allows us to use measurements from the primary detector up to 4.2 $\times$ 10$^{6}$ counts/s. Right: Count rate of both detectors in mass channel 2 as a function of latitude for orbit 290. Closest approach to Saturn, where the signal is highest, occurs near -5$^{\circ}$ latitude. As the spacecraft approaches Saturn, the primary detector (black) saturates which leads to a signal decay, while the lower signal secondary detector (blue) does not. We are able to combine measurements from both detectors and determine a corrected count rate to be used to determine a proper H$_2$ density in Saturn's atmosphere (red).}
\label{fig:detector}
\end{figure}

The instrument's dual detector system includes a high gain primary detector which is known to saturate near closest approach in mass channels relevant to the most abundant species in the atmosphere. This leads to signal decay at closest approach as seen in Figure \ref{fig:detector}. When this occurs, counts from the secondary detector, albeit with a lower signal-to-noise ratio, can be utilized to obtain an accurate determination of the density profile. In order to take advantage of the primary detector's much higher signal-to-noise ratio, we use a nonlinear conversion technique between the primary and secondary detectors as detailed in \citeA{cui12}. In doing so, we are able to utilize slightly saturated count rates from the primary detector, significantly reducing the amount of noise in the density profile. Saturation of the primary detector is a species dependent process. At Titan, detector saturation occurred in mass channels associated with the atmosphere's major constituents, N$_2$ and CH$_4$, at \textit{m/z} 14, 15, 16, 28, and 29 amu. At Saturn, the only channel that experiences saturation is \textit{m/z} 2 amu, which tracks H$_2$ in the atmosphere. This saturation can be seen in Figure \ref{fig:detector}, where the count rate from the primary detector in mass channel 2 is plotted against the count rate from the secondary detector for all of the final orbits. At lower counts the detectors are linearly correlated but as the count rate of the primary detector increases, the signal begins to decay and the detectors lose their linear correlation. The relationship between the detectors can be described using an empirical equation, as described in \citeA{cui12}:

\begin{linenomath*}
\begin{equation}
C_2 = a_0 C_1 \exp \{ \tan[(a_1 C_1)^{a_2}] \},
\end{equation}
\label{eq:detector_relationship}
\end{linenomath*}

where $a_0$, $a_1$, and $a_2$ are free parameters constrained by the data. The free parameters for each orbit are listed in Table \ref{tab:free_parameters}. Characterizing this relationship allows us to use slightly saturated counts from the primary detector up to 4.2 $\times$ 10$^{6}$ counts/s. At this point, the empirical relationship no longer traces the primary detector's signal decay and the correction is no longer applicable, prompting the use of counts from the secondary counter. However, an instantaneous switch from the primary to secondary detector at 4.2 $\times$ 10$^{6}$ counts/s could introduce discontinuities in the derived density profiles, which in turn would affect the retrieved temperature profile. To remove this effect, we introduce continuously varying weighting functions to calculate densities in the transition region as detailed in \citeA{cui12}. 

\begin{table}
\caption{Free parameters used in the empirical relationship between the count rates from the primary and secondary detectors for mass channel 2 (H$_2$.)}
\centering
\begin{tabular}{c c c c}
\hline
Orbit & $a_0$ & $a_1$ & $a_2$ \\
Number & & & \\ 
\hline
288 &  3.291 $\times$ 10$^{-4}$ & 1.689 $\times$ 10$^{-7}$ & 2.556\\
290 & 3.975 $\times$ 10$^{-4}$ & 1.506 $\times$ 10$^{-7}$  & 3.223\\
 291 &  3.718 $\times$ 10$^{-4}$ & 1.362 $\times$ 10$^{-7}$  & 3.061\\
 292 & 3.632 $\times$ 10$^{-4}$ & 1.351 $\times$ 10$^{-7}$  & 3.117\\
 293 & 3.528 $\times$ 10$^{-4}$ & 1.507 $\times$ 10$^{-7}$  & 7.020 \\
 All & 3.378 $\times$ 10$^{-4}$ & 1.373 $\times$ 10$^{-7}$  & 2.620\\
\hline 
\end{tabular}\\
\label{tab:free_parameters}
\end{table}

The INMS flight unit (FU) was calibrated with neutral species relevant to Titan's atmosphere before launch and additional calibration of other species continued after launch using the identical engineering unit (EU). Thus, a calibration database to compare known fragmentation patterns to observed measurements exists. Due to response differences between the FU and EU, calibration of the peak sensitivity of each species must be performed in order to utilize EU calibration measurements to understand observations using the FU. \citeA{cui09a} developed an algorithm to characterize these sensitivity differences based on measurements of species that were calibrated using both units and we utilize these results here since many of these molecules are also relevant to Saturn's upper atmosphere. Additionally, the ram pressure of the inflowing sample in CSN mode leads to a density enhancement in the instrument that varies as a function of molecular mass, angle of attack of the instrument, temperature of the ambient gas, and speed of the spacecraft. This ram enhancement factor was previously characterized in \citeA{cui09a} for Titan flybys, and we use the same approach to correct for the ram enhancement factor here. Corrections for contamination from thruster firings of the spacecraft, which occasionally affect the counts in mass channel 2 during Titan flybys, are not done here since thrusters were not used near closest approach for these final orbits.

Additionally, Saturn's high rotation rate and significantly oblate shape invalidates the common assumption that atmospheric variations are purely radial. Instead, we assume here that Saturn's atmospheric properties vary with gravitational potential, $\phi$, and we use $\phi$ as the vertical coordinate in our analysis. This modification includes adopting the updated gravitational potential for Saturn found in \citeA{anderson07}. This process is detailed in the supporting information of \citeA{yelle18}. 

Although the instrument records data before and after closest approach to the planet, our analysis utilizes solely inbound data. INMS has a well-documented behavior where certain species entering the instrument adsorb to the chamber's walls. This adsorption can lead to wall chemistry within the instrument and/or desorption at a later time (see e.g., \citeA{cui08, vuitton08, cui09a}). This effect is observed near closest approach to the planet, when the number density of molecules is highest, and predominantly affects outbound measurements which leads to an artificial outbound density enhancement for certain species. For this reason, we do not utilize these measurements here and focus on direct inbound measurements. 

\begin{figure}
\noindent\includegraphics[width=\textwidth]{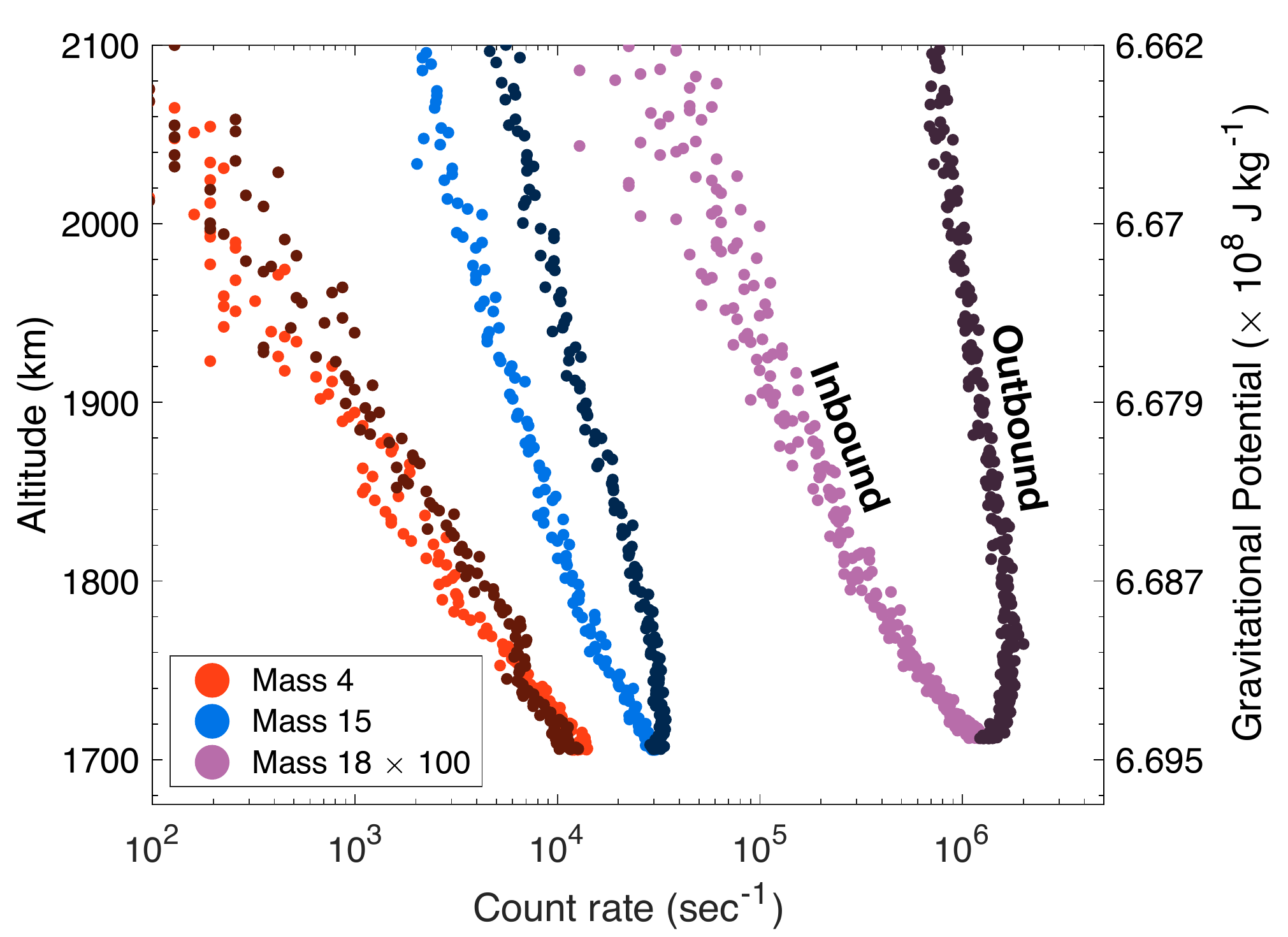}
 \caption{Count rate of mass channels 4, 15, and 18 as a function of altitude and gravitational potential from orbit 290. The lighter shade for each mass channel represents the inbound profile and the darker shade represents the outbound profile. The inbound and outbound profiles of mass channel 4 (He) are nearly identical since this species does not adsorb to the instrument's chamber walls or participate in wall chemistry. Mass channel 18 (H$_2$O) is known to be affected by wall adsorption and chemistry in the instrument, which is the reason for the significant inbound/outbound asymmetry. Mass channel 15 is a combination of signal from CH$_4$ and NH$_3$. Since CH$_4$ is not affected by wall adsorption in the instrument, this asymmetry indicates that NH$_3$ must be contributing to the signal.}
 \label{fig:inbound_outbound}
\end{figure}

While wall adsorption leads to difficulties in properly interpreting the outbound measurements for some mass channels, this effect is actually valuable in determining what species might be present in the measurements. Only certain species are affected by wall adsorption and subsequent chemistry within the instrument. Inert species and CH$_4$ are known not to contribute to wall adsorption (see e.g., \citeA{cui08, cui09a}), whereas other species, such as H$_2$O and NH$_3$, exhibit significant effects due to wall adsorption. Using this knowledge, it can be deduced whether certain species are present in the measurements based solely on the inbound/outbound asymmetry of relevant mass channels. An example of this is seen in Figure \ref{fig:inbound_outbound}. Mass channel 4, which tracks He (an inert species), exhibits a symmetric profile before and after closest approach. On the other hand, mass channel 18, which tracks H$_2$O, exhibits a significant asymmetric distribution before and after closest approach indicative of wall effects for this species. Mass channel 15 is primarily a combination of signal from fragments associated with CH$_4$ and/or NH$_3$. The asymmetric distribution implies that the signal measured in this mass channel must have some contribution from NH$_3$ since the signal from a lack of NH$_3$ would have had no asymmetry at closest approach. In this figure, the left axis of altitude is the height above Saturn's 1-bar pressure level that corresponds to the gravitational potential values seen on the right axis. The profile of inbound measurements for mass channels 15 and 18 suggests that measurements in these channels directly before closest approach (below $\phi$ of approximately 6.69 $\times$ 10$^{8}$ J kg$^{-1}$ (1750 km) could also be slightly affected by a density enhancement for orbits 288 to 292. Proper correction for this enhancement would require extensive knowledge of how each species interacts within the chamber and subsequent modeling to correct for this asymmetry and is beyond the scope of this paper. 

\subsection{Mass spectral deconvolution}

\begin{figure}
\noindent\includegraphics[width=\textwidth]{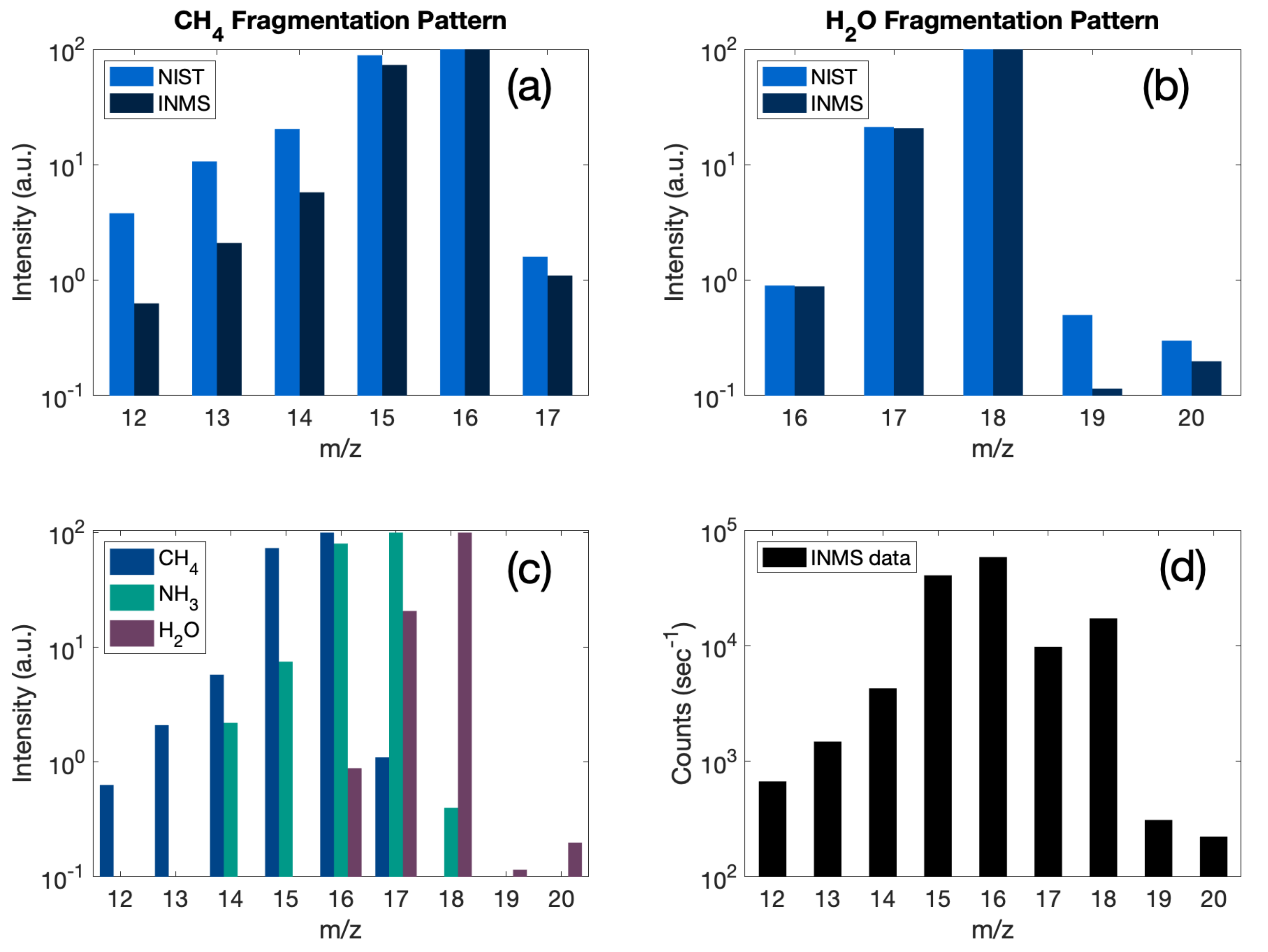}
 \caption{(a and b) Comparison of INMS and NIST calibration data for CH$_4$ (a) and H$_2$O (b). The differences in fragmentation peak intensities, along with a lack of INMS calibration data for some species, complicate analysis of the spectra returned by INMS. (c) Overlapping fragmentation patterns of CH$_4$, H$_2$O, and NH$_3$ which all contribute to the INMS signal in the region m/z = 12 - 20 amu. CH$_4$ and H$_2$O data are from INMS calibration measurements. NH$_3$ data are from the NIST spectral library. (d) Signal from INMS from orbit 290 \add{extracted between $\phi$ of 6.69 and 6.66 $\times$ 10$^{8}$ J kg$^{-1}$ (1700 to 2050 km).}}
 \label{fig:massfrag}
\end{figure}

Species identification and density determination from INMS measurements is complicated by the fact that multiple species contribute to the signal of individual mass channels, creating a complex combination of mass peaks associated with a mix of the fragmentation patterns of the species within the sample. An accurate density determination of species within the atmosphere must begin by first determining the relative contribution of each species to each mass channel. This is done by combining the peak intensities of the fragmentation patterns of each species from calibration data weighted by a relative contribution from each species that results in the best fit to the measured mass spectrum. Thus, one must solve a system of linear equations:
\begin{linenomath*}
\begin{equation}
I_{i} = \sum_{j = 1}^{n}F_{i,j}N_{j}
\end{equation}
\label{eq:mass_spec}
\end{linenomath*}
The relative intensities of the fragmentation peaks produced by a certain species are highly dependent on the instrument used, thus it's crucial to have an accurate instrumental fragmentation pattern database to compare to observations returned by the spacecraft. As stated above, we focus here on the region of the mass spectrum from 12 to 20 amu and the primary species contributing to this region: CH$_4$, H$_2$O, and NH$_3$. \add{An example of the signal measured by INMS in this region is shown in Figure 3d for orbit 290.} Figure \ref{fig:massfrag}c compares the overlapping fragmentation patterns of these three species from calibration data. Although INMS was calibrated for many species relevant to both Titan and Saturn, unfortunately the instrument was never calibrated for NH$_3$. As a consequence, calibration data from the National Institute of Standards and Technology (NIST) mass spectral library must be utilized for NH$_3$. Figures \ref{fig:massfrag}a and b compare the INMS calibration data of CH$_4$ and H$_2$O to data from NIST. Although NIST calibrations provide an adequate estimate on what to expect during the flight instrument's performance, there are significant deviations in fragmentation peak intensity in certain mass channels for both species. Furthermore, the instrument's calibration on Earth was performed in an environment very different from that of Saturn, which could lead to discrepancies between the fragmentation patterns found in the calibration database and the actual measurements returned from the instrument. Deviations stemming from the aging of the instrument, which was launched in 1997, could also affect the instrument's performance over time and lead to further discrepancies between calibration values and returned measurements. Although a calibration database is of utmost importance, an accurate and complete database of fragmentation patterns relevant to this study does not exist. 

In order to overcome the challenges brought about by the calibration data from INMS, our mass spectral deconvolution algorithm employs a Monte-Carlo based approach to handle the uncertainty in fragmentation peak intensities of each species \cite{gautier19}.  \change{The Monte-Carlo randomization is applied to the intensity of each fragmentation peak for each species in the database, not to the location of each peak in the spectrum.}{The Monte-Carlo randomization is applied only to the intensity (y-axis) of each fragmentation peak for each species in the database, not to the \textit{m/z} ratio (x-axis) of the fragmentation peaks.} We allow for $\pm$15\% variation in individual peak intensities from the original database, which is a combination of INMS calibration data and NIST values when INMS data is unavailable. This process is done 500,000 times, thus creating 500,000 individual fragmentation databases that are then used to decompose the measured mass spectrum. The output of the deconvolution includes the relative abundances of each species in the database based on the randomized fragmentation pattern database that was input into the model. We keep the best-fitting 2\% (best-fitting 10,000 mass spectral fits) of these simulations for analysis based on the outputs with the minimal residuals to the data. This allows us in the end to retrieve a statistical solution to our issue, providing the most probable concentration for a given species as well as a probability density function (PDF) in order to quantify the variation in a species' concentration throughout the best-fitting simulations saved for analysis. Although we allow the peak intensities to vary by 15\%, the main peaks for the best-fitting simulations saved for analysis vary by only a few percent and the less significant peaks typically vary no more than $\sim$10\%. 

\begin{figure}
\noindent\includegraphics[width=\textwidth]{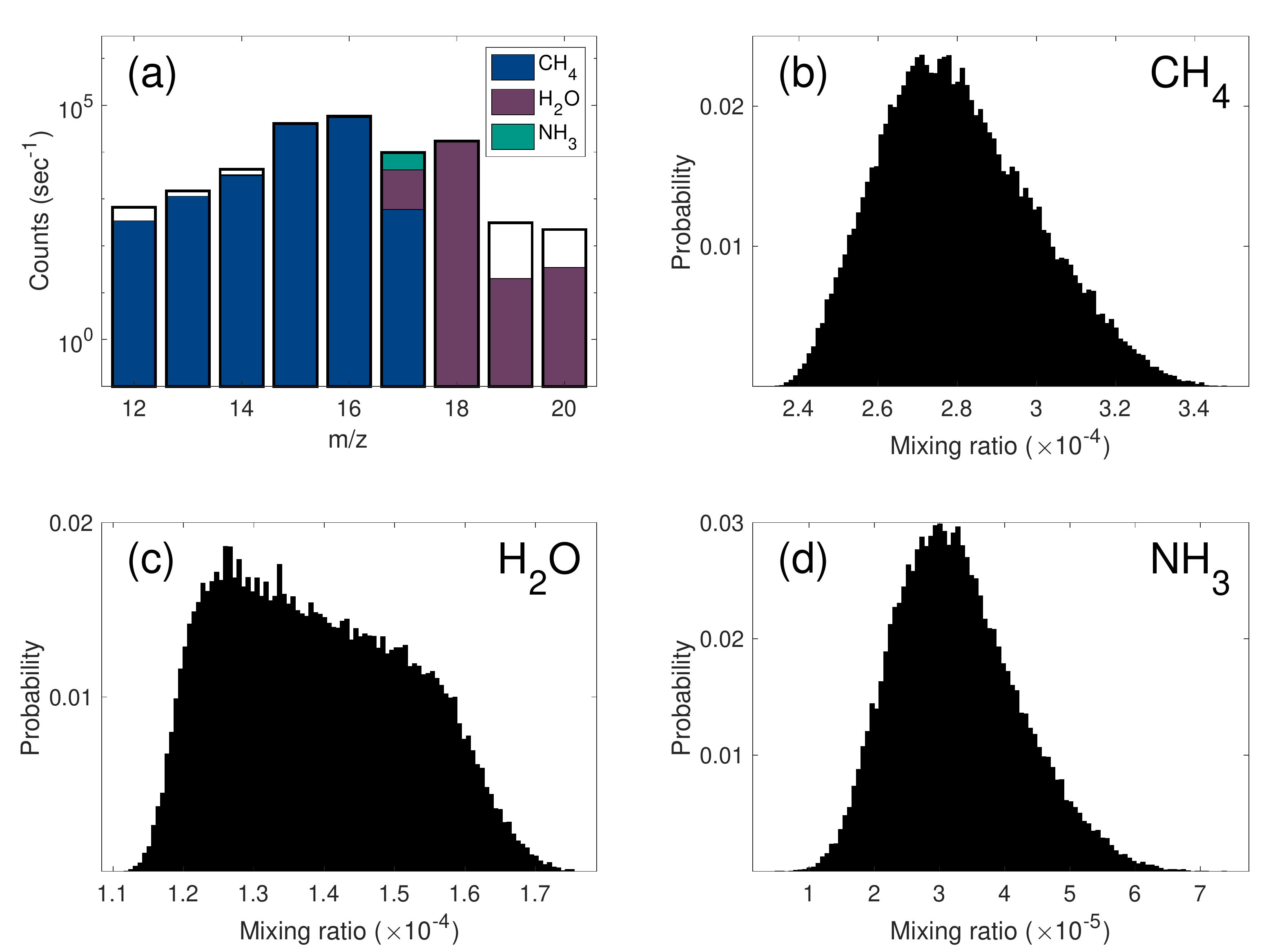}
\caption{ (a) Example of mass spectrum deconvolution result using our Monte-Carlo approach for orbit 290. Black outline bars represent the measured INMS spectrum. Colored segments represent the contribution of each species as calculated by the model. Contributions shown here are the average contribution of the best-fitting 2\% of 500,000 simulations. (b--d) Probability densities of the mixing ratio of each species in the sample as retrieved by the model for the best-fitting 2\% of simulations.}
\label{fig:mzfit}
\end{figure}

In solving for the relative concentration of each species, we are able to determine the mixing ratio for these species from our model. An example of a mass spectral fit along with PDFs is shown in Fig. 4 for orbit 290 with an averaged mass spectrum from a gravitational potential height of \change{6.65 $\times$ 10$^{8}$ J kg$^{-1}$  down to closest approach.}{6.69 to 6.66 $\times$ 10$^{8}$ J kg$^{-1}$.}\note{I noticed while revising the manuscript that this statement was not properly updated after the methods had slightly changed. The region of the atmosphere that was used to created the averaged mass spectra has been updated, and now matches the rest of the manuscript (i.e., fig 3d and fig 6)} The black outline bars in Figure \ref{fig:mzfit} show the measured mass spectrum and the colored bars represent the modeled contribution of each molecule to the mass channel. For this study, we use H$_2$, He, CH$_4$, H$_2$O, and NH$_3$ to fit the mass spectrum up to 20 amu. There is no signal detected from 5 to 11 amu. The signal below 5 amu is predominantly from H$_2$ and He, and above 11 amu is a combination of CH$_4$, H$_2$O, and NH$_3$. A detailed analysis for H$_2$ and He was published previously in \citeA{yelle18}. These species will be discussed here only for comparison and are included in our modeling in order to retrieve accurate mixing ratio information since these are the two major constituents of Saturn's atmosphere. We assume that the region from 12 to 20 amu is contributed only by CH$_4$, H$_2$O, and NH$_3$. In reality, it is likely that fragments of other molecules with higher masses also contribute to these mass channels. However, higher mass species will not dissociate into fragments that contribute significantly to the predominant peaks associated with CH$_4$, H$_2$O, and NH$_3$ (16, 17, and 18 amu), so the small contributions from the fragments of higher mass species are not included in this study. These fragments will mostly contribute to mass channels 12 and 14 amu (corresponding to signal from ionized carbon and nitrogen, respectively), which are consistently underfit by our model. As a consequence, the results presented here should be considered upper limits for the mixing ratios of CH$_4$, H$_2$O, and NH$_3$. Future work will address the entirety of the mass spectrum (up to 99 amu), which could result in a slight revision to the mixing ratio values presented here. In any case, the modeled spectra provide reasonable fits to the data, with the major peaks from 15 to 18 amu being reproduced by the model very well. Mass channels 19 and 20, whose signal has contributions from isotopes of H$_2$O, are underfit by our current modeling efforts. The signal at 19 amu is significantly higher than expected and could be a consequence of contamination from filament desorption (see e.g., \citeA{perry10, perry15}), however this source of background contamination is poorly quantified. Mass channel 20 could have contributions from other molecules at higher masses, including argon which contributes to the signal as Ar$^{2+}$. Nonetheless, accurate measurements of isotopes for these species are beyond the scope of this paper and will be addressed in upcoming work.

\subsection{Density determination}

After correcting the raw INMS counts as described above, we are able to utilize the mass spectral deconvolution algorithm to derive number density profiles for the species. In order to do so, the count rates for each mass channel are divided into bins with a width of $\phi$ = 0.01 J kg$^{-1}$ and the deconvolution is performed for each separate gravitational potential bin. The average output of the best-fitting 2\% of simulations is then used to determine the relative contribution of each species to its dominant mass channel. This is done using mass channels 16, 17, and 18 amu, which are the molecular peaks of CH$_4$, H$_2$O, and NH$_3$, respectively, and have very high signal. CH$_4$ and H$_2$O both have peaks which are relatively free of significant contamination by other species: 12 and 13 amu for CH$_4$, and 18 to 20 amu for H$_2$O. This, along with comparison to the peak intensities within the region where there is contribution from other species, provides convincing evidence that these species are in fact present in the measured mass spectrum. The NH$_3$ contribution, on the other hand, must be constrained by mass channel 17, which is the species' parent peak but has significant contribution from both CH$_4$ and H$_2$O. Thus, NH$_3$ detection and quantification relies more heavily on our modeling efforts than that of CH$_4$ and H$_2$O. However, as stated previously, there is evidence that NH$_3$ is present within the spectrum due to the asymmetry of the measurements around closest approach as shown in Figure \ref{fig:inbound_outbound}.

We utilize counts from the instrument as long as the signal-to-noise ratio is sufficient for analysis. For He, H$_2$O, and NH$_3$, the loss of signal occurs at a higher $\phi$ (lower altitude) than CH$_4$ and H$_2$ due to the lower abundance of these three species in the region. Thus, we obtain density profiles for H$_2$ and CH$_4$ that extend to a lower $\phi$ (higher altitude) from closest approach than the other species. After running the Monte Carlo algorithm described above, we determine the density profiles by weighting the corrected count rate from the species' main mass channel with the relative contribution of that species returned by the model for each $\phi$ bin. Results using this method are shown in Fig.~5. Running independent simulations in each $\phi$ bin provides us with reasonable results for each species. The average output of each $\phi$ bin from the best-fitting 2\% of simulations is then combined to create a density profile, which produces smooth and consistent profiles for each species. This consistency demonstrates the power of our new method in helping to determine the contribution of each species when the available calibration data for an instrument is insufficient. Errors associated with these density profiles are a combination of 1$\sigma$ uncertainties due to counting statistics from the data and 1$\sigma$ standard deviation from the modeling results. H$_2$ and He have a higher signal-to-noise ratio in their dominant mass channels and are better constrained by the model, leading to small errors associated with these species. They are also well constrained by the model since there is no interference from other species in their relevant mass channels. Measurements for H$_2$O and NH$_3$ have a lower signal and these species have a greater spread in the modeling results, leading to slightly larger error bars for these species. \add{The retrieved errors associated with the density profiles in Fig.~5 for all molecules but H$_2$ range from a few percent near closest approach, where signal is highest, and increase to approximately 50\% for the uppermost measurements where signal is much lower. Errors in the H$_2$ density profile are no more than 10\% for the region of the atmosphere analyzed here.} 

\section{Results and Discussion}
\subsection{Density profiles and mixing ratios}

\begin{figure}
\noindent\includegraphics[width=\textwidth]{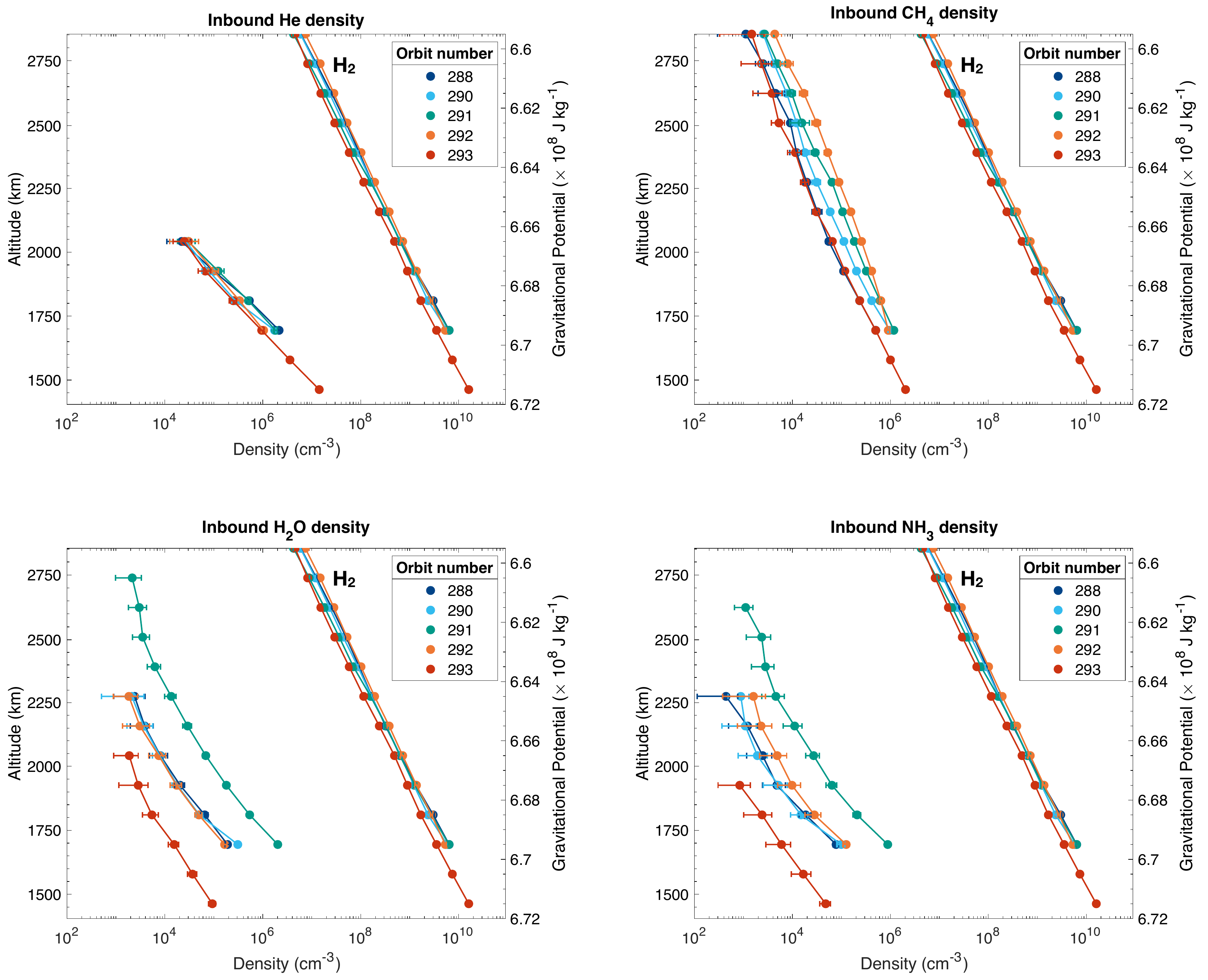}
\caption{Inbound density profiles of He, CH$_4$, H$_2$O, and NH$_3$ versus altitude and gravitational potential for all five orbits analyzed. The density profiles of H$_2$ are also included in each figure for comparison. Density profiles are constructed by averaging INMS measurements in gravitational potential bins of 0.01 J kg$^{-1}$, then running a mass spectral deconvolution for each individual bin. Error bars are a combination of 1$\sigma$ uncertainties from counting statistics and 1$\sigma$ uncertainties from the mass spectral deconvolution.}
\label{fig:density}
\end{figure}

Density profiles of H$_2$, He, CH$_4$, H$_2$O, and NH$_3$ from Cassini's last orbits are shown in Figure \ref{fig:density}. Although these results agree with and are an extension of the work presented in \citeA{yelle18}, we obtain the density profiles of H$_2$, He, and CH$_4$ using slightly different methods. In \citeA{yelle18}, we obtain density profiles of these species by using the corrected measurements in the dominant mass channels for each species (\textit{m/z} 2, 4, and 16 amu, respectively) without fitting the mass spectrum. This method is reasonable since there is very little interference from other species in these three mass channels. In order to expand our analysis to H$_2$O and NH$_3$, we determine density profiles by binning the data and employing the mass spectral deconvolution as described above. Our new method does not modify the results of H$_2$ and He since there is no overlapping signal in their dominant mass channels. The results of CH$_4$ are changed by about 2\% on average, which is well within the errors associated with the measurements. 

We obtain density profiles for CH$_4$ from closest approach for each orbit up to $\phi$ of approximately 6.59 $\times$ 10$^{8}$ J kg$^{-1}$ (2850 km) and for He from closest approach up to $\phi$ of approximately 6.67 $\times$ 10$^{8}$ J kg$^{-1}$ (2050 km). H$_2$O and NH$_3$ show greater variability in their density profiles from orbit to orbit, leading to differences in signal and consequently differences in the extent of their retrieved density profiles. The profiles presented here provide further evidence from INMS for an external source providing Saturn's upper atmosphere with a variety of unexpected species, as previously reported in \citeA{yelle18, waite18, perry18, miller20}. The density of species as heavy as CH$_4$, H$_2$O, and NH$_3$ should decrease rapidly above Saturn's homopause as a consequence of diffusive separation. However, the scale heights of these species and their presence in this region of the atmosphere are inconsistent with these species being native to Saturn's interior and emerging via upward diffusion. He, on the other hand, decreases rapidly above the homopause and follows the expected scale height and density profile of a molecule of its mass in a H$_2$-dominated atmosphere.

\begin{figure}
\noindent\includegraphics[width=\textwidth]{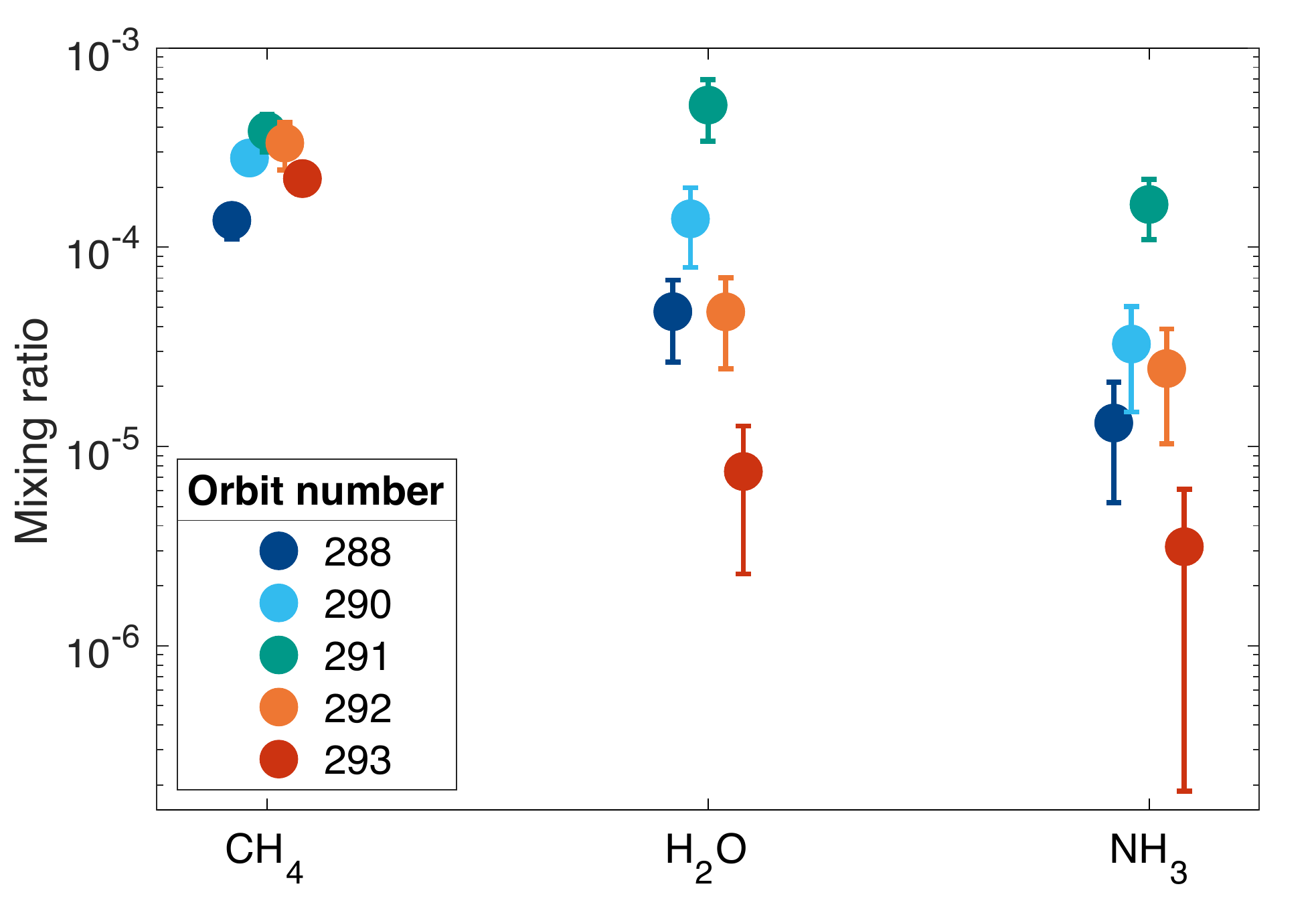}
\caption{Averaged mixing ratios of CH$_4$, H$_2$O, and NH$_3$. We use measurements taken between $\phi$ of 6.69 and 6.66 $\times$ 10$^{8}$ J kg$^{-1}$ (1700 to 2050 km), where reliable data for all orbits exist, to create the integrated spectra used to determine mixing ratios. Variability among orbits for these volatile species is evident, especially for orbit 293 (atmospheric entry) which measured a different region of Saturn at closest approach and shows a large depletion in H$_2$O and NH$_3$. Error bars are a combination of 1$\sigma$ uncertainties from counting statistics and 1$\sigma$ uncertainties from the mass spectral deconvolution.}
\label{fig:MR}
\end{figure}

Figure \ref{fig:MR} shows the average mixing ratio for each species for each orbit. In order to compare the orbits, we determine these mixing ratios using an averaged mass spectrum that incorporates the region of the atmosphere where all five orbits have reliable data. This corresponds to the region of the atmosphere of approximately 6.69 to 6.66 $\times$ 10$^{8}$ J kg$^{-1}$ (1700 to 2050 km). One notable observation is the variability of these external species, and especially the depletion of H$_2$O and NH$_3$ during orbit 293 (atmospheric entry). All orbits aside from orbit 293 measured Saturn's atmosphere at similar conditions with closest approach to the planet around 5$^{\circ}$ S (see Table \ref{tab:orbitalinfo}). Orbit 293, however, entered Saturn's atmosphere around 9$^{\circ}$ N before crossing the equatorial ring plane. The difference in the region sampled for orbit 293 could be the explanation for depletion in H$_2$O and NH$_3$ during atmospheric entry. 

\begin{figure}
\noindent\includegraphics[width=\textwidth]{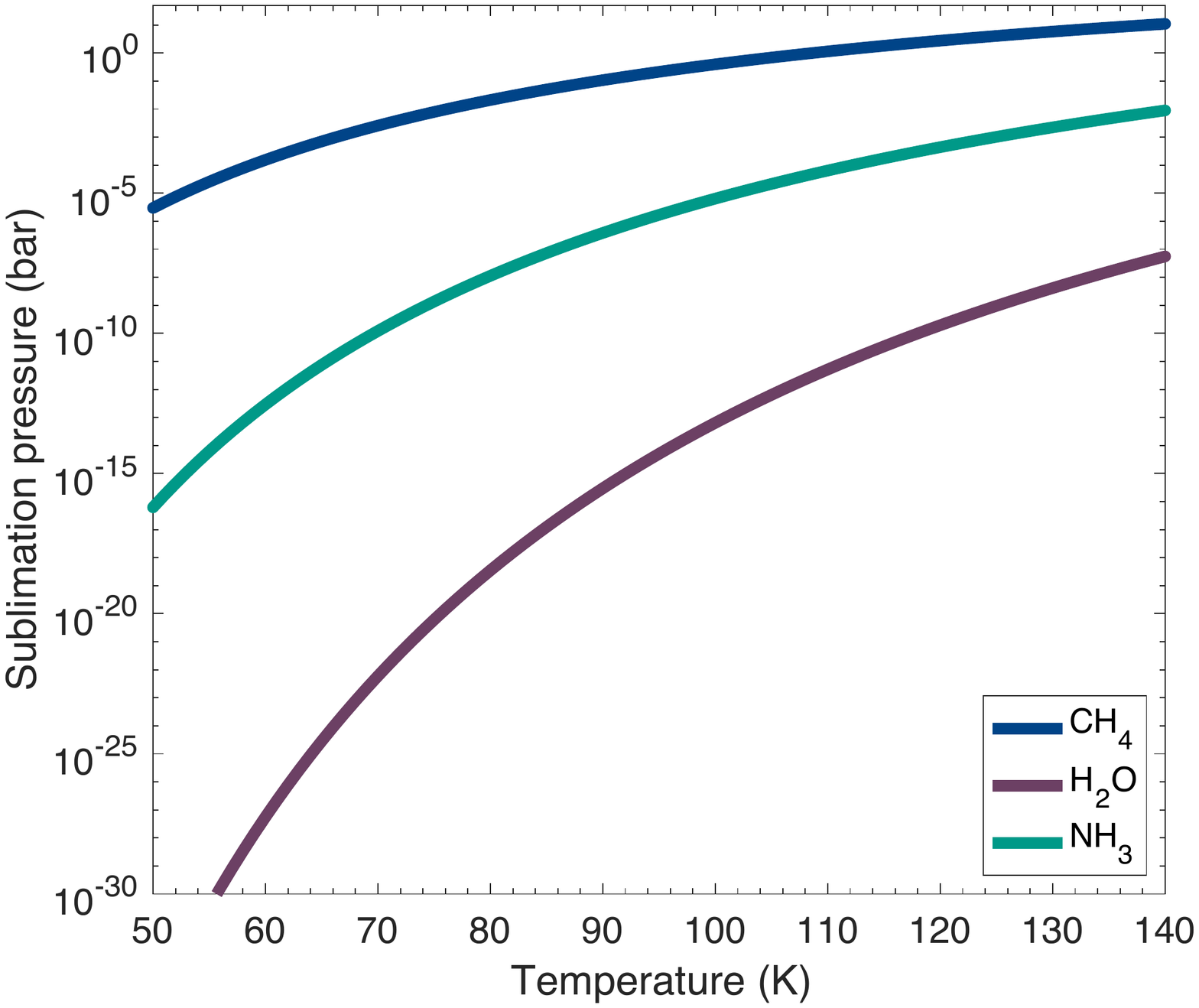}
\caption{Sublimation pressure curves for CH$_4$, H$_2$O, and NH$_3$ \cite{fray09} at temperatures relevant to Saturn's rings. }
\label{fig:SVP}
\end{figure}

The mixing ratio of He shows very little variation and is around 3.0 $\pm$0.2 $\times$ 10$^{-4}$ for this region of the atmosphere. CH$_4$ shows more variability than He, but measurements are roughly consistent among orbits, ranging from 1.4 to 3.8 $\times$ 10$^{8}$. The large variability of H$_2$O and NH$_3$ among orbits, as well as the prevalence of CH$_4$ as a dominant external species, are surprising and unexpected, but potential explanations for these results do exist. First, as noted in \citeA{yelle18}, the prevalence of CH$_4$ over H$_2$O and NH$_3$ could be tied to the volatility of these species. Their sublimation pressures, shown in Figure \ref{fig:SVP}, vary by orders of magnitude at temperatures relevant to Saturn's inner rings of $\sim$80 to 115 K \cite{filacchione14, tiscareno19}. Saturn's rings are constantly exposed to magnetospheric plasma, cosmic ray impacts, micrometeorite bombardment, and other high energy phenomena, resulting in the ejection of surface ring material which liberates gas molecules from the rings. If these species are present in the rings and are liberated from ice particles in this way, then it is more likely for emitted H$_2$O and NH$_3$ molecules to recondense back onto ring particles while CH$_4$ is preferentially lost into Saturn's atmosphere.

It is also possible that H$_2$O and NH$_3$ are entering Saturn's atmosphere from the rings in a charged form (i.e., H$_3$O$^+$ and NH$_4$$^+$), which would elude detection by INMS. H$_2$O and NH$_3$ have much higher proton affinities (691.0 and 853.6 kJ mol$^{-1}$, respectively) than CH$_4$ (543.5 kJ mol$^{-1}$) \cite{hunter98}, making them more susceptible to proton transfer. If H$_2$O and NH$_3$ are more easily protonated after liberation from ring particles, they would not be detected by INMS since the ion mode of the instrument was limited to masses below 8 amu during these orbits due to the speed of the spacecraft \cite{cravens18, moore18}. It is also possible that these charged molecules are transported away from the equatorial region via magnetic field lines and deposited into Saturn's atmosphere at higher latitudes that were not observed by INMS. Indeed, \citeA{odon17} recently reported evidence that charged water was entering Saturn's midlatitude region from the rings via magnetic field lines. An improved understanding of these potential transport processes will require continued monitoring from ground and space based observations and additional in situ measurements from future missions.

\subsection{Flux and mass deposition rate calculations}

To quantify the influx of material from the rings, we use a 1-D model to understand the exospheric temperature and external flux in the upper atmosphere. This model is described extensively in the supporting information of \citeA{yelle18}, who use the model to provide a detailed analysis of the CH$_4$ influx into Saturn's atmosphere. We follow the same method here with the addition of H$_2$O and NH$_3$. Briefly, we assume hydrostatic equilibrium and solve the standard diffusion equation with $\phi$ as the vertical coordinate. The mixing ratio of a species is given by
\begin{linenomath*}
\begin{equation}
X_{i}(\phi) = X_{i}(\phi_\circ)\exp\int_{\phi_\circ}^{\phi}\mathrm{d}{\phi}'\frac{D_i}{D_i + K}\frac{m_i - m_a}{RT} - \int_{\phi_\circ}^{\phi}\mathrm{d}{\phi}'\frac{F_i}{gN_a(D_i + K)}\exp(\int_{\phi_\circ}^{\phi}\mathrm{d}{\phi}''\frac{D_i}{D_i + K}\frac{m_i - m_a}{RT})
\label{eq:diffusion}
\end{equation}
\end{linenomath*}
where $X_i$ is the mixing ratio of the minor constituent, $D_i$ is the molecular diffusion coefficient, $K$ is the eddy diffusion coefficient, $m_i$ is the molecular mass of the minor constituent, $m_a$ is the average molecular mass of the atmosphere, $R$ is the gas constant for H$_2$, $T$ is the temperature, $g$ is the magnitude of the gravitational acceleration, $N_a$ is the density of H$_2$, and $F_i$ is the flux of the minor constituent. The first term of this equation represents molecular diffusion within the atmosphere and the second term describes the vertical distribution of a molecule with a non-zero external flux into the atmosphere. 

We determine the temperature using the H$_2$ density profile for each orbit. For the plunge, which probes deeper into the atmosphere, we determine the temperature profile using a Bates profile for a thermosphere \cite{bates51}. The other orbits probe higher in the atmosphere and we determine the temperature by fitting an isothermal model to the H$_2$ density profiles. The temperature results are found in Table 3 and were originally published in \citeA{yelle18}, where further discussion and interpretation is presented. We also adopt the eddy diffusion profile from \citeA{yelle18}, which has a constant value in the uppermost region of the atmosphere of K$_\infty$ = 1.4 $\times$ 10$^{4}$ m$^2$s$^{-1}$, although a wide variety of K profiles can be used to fit the data. The molecular diffusion coefficients for He and CH$_4$ in H$_2$ are taken from \citeA{mason70}. We calculate the molecular diffusion coefficients for H$_2$O and NH$_3$ in H$_2$ with the theoretical approach outlined in \citeA{hirschfelder54} based on the Lennard-Jones potential. Molecular diffusion coefficients for these species at $\phi$ of 6.69 $\times$ 10$^{8}$ J kg$^{-1}$ are 1.3 $\times$ 10$^{6}$ m$^2$s$^{-1}$ for He, 6.2 $\times$ 10$^{5}$ m$^2$s$^{-1}$ for CH$_4$, 8.0$\times$ 10$^{5}$ m$^2$s$^{-1}$ for H$_2$O, and 8.4 $\times$ 10$^{5}$ m$^2$s$^{-1}$ for NH$_3$.

Helium is native to the planet and exhibits a mixing ratio decreasing with altitude in accordance with a species following diffusive separation in a well-mixed atmosphere. Thus, our model needs no external flux (F$_i$ = 0) to properly fit the He measurements. On the other hand, H$_2$O, CH$_4$, and NH$_3$ all exhibit roughly constant mixing ratios in the upper atmosphere, which is expected for a minor constituent entering the atmosphere externally \cite{connerney84}. To determine the external flux we utilize inbound measurements taken before the slight density enhancement observed for H$_2$O and NH$_3$, which excludes measurements taken below $\phi$ of 6.69 $\times$ 10$^{8}$ J kg$^{-1}$ (1750 km) for orbits 288 to 292. Measurements can be fit using a constant external flux and results for all orbits are found in Table 3. 

\begin{table}
\caption{Temperature, flux, and mass deposition rate results.}
\centering
\begin{tabular}{c c c c c}
\hline
Orbit & Temperature & Molecule & Flux & Mass Deposition \\
Number & (K) & & ($\times$ 10$^{12}$ m$^{-2}$s$^{-1}$) & Rate ($\times$ 10$^{2}$ kg s$^{-1}$) \\ 
\hline
288 & 368.8 $\pm$ 1.1 & CH$_4$ & 7.0 $\pm$ 1.3 & 12.5 $\pm$ 2.3\\
& & H$_2$O & 1.6 $\pm$ 0.7 & 3.1 $\pm$ 1.4\\
& & NH$_3$ & 0.5 $\pm$ 0.3 & 1.0 $\pm$ 0.5 \\
& & Total & 9.0 $\pm$ 1.5& 16.6 $\pm$ 2.7\\
 \hline 
290 & 363.7 $\pm$ 1.0 & CH$_4$ & 13.9  $\pm$ 1.6 & 25.0 $\pm$ 2.8\\
& & H$_2$O & 1.5 $\pm$ 0.6 & 2.9 $\pm$ 1.3\\
& & NH$_3$ & 0.5 $\pm$ 0.2 & 1.0 $\pm$ 0.5\\
& & Total & 15.9 $\pm$ 1.7 & 29.0 $\pm$ 3.1\\
\hline  
 291 & 339.6 $\pm$ 1.2 & CH$_4$ & 24.7  $\pm$ 5.1 & 44.3 $\pm$ 9.1\\
 && H$_2$O & 13.7  $\pm$ 4.5 & 27.6 $\pm$ 9.0\\
 && NH$_3$ & 5.6 $\pm$ 1.8 & 10.7 $\pm$ 3.3\\
 && Total & 44.0 $\pm$ 7.0 & 82.6 $\pm$ 13.2\\
 \hline 
 292 & 372.1 $\pm$ 1.0 & CH$_4$ & 25.7  $\pm$ 6.8 & 46.0 $\pm$ 1.2\\
 && H$_2$O & 1.2  $\pm$ 0.5 & 2.3 $\pm$ 1.1 \\
 && NH$_3$ & 1.0 $\pm$ 0.4 & 1.8 $\pm$ 0.7\\
 && Total & 27.8 $\pm$ 6.8 & 50.2 $\pm$ 12.3\\
 \hline 
 293 & 351.1 $\pm$ 1.2 & CH$_4$ & 12.2  $\pm$ 1.6 & 19.8 $\pm$ 2.8\\
 && H$_2$O & 0.4  $\pm$ 0.4 & 0.8 $\pm$ 0.8\\
 && NH$_3$ & 0.1 $\pm$ 0.1 & 0.3 $\pm$ 0.2\\
 && Total & 11.7 $\pm$ 1.6 & 20.8 $\pm$ 2.9\\
\hline 
\end{tabular}\\
\label{tab:results}
\end{table}

The variability in the influx of external material measured among orbits, especially for H$_2$O and NH$_3$, leads to large fluctuations in the observed amount of material deposited into the atmosphere. We quantify the mass deposition rate (MDR) for each species into the atmosphere using the equation
\begin{linenomath*}
 \begin{equation}
MDR = F_i m_i \times 2\pi r_{Saturn}^2\theta,
\end{equation}
\end{linenomath*}
where $F_i$ is the flux of molecule $i$, $m_i$ is the molecular mass of molecule $i$, and the second term represents the surface area of Saturn that is affected by the deposition of external material. We use a latitudinal width ($\theta$) for the influx region of 16$ ^{\circ}$. This width corresponds to the location at which we observe a decrease in signal of the minor constituents, which occurs at approximately $\pm$8$ ^{\circ}$ from the ring plane. It is possible that the affected region of Saturn is larger than this, however a larger width cannot be deduced from our measurements due to the loss of signal. 

Mass deposition rates for all three species are found in Table 3. Although previous studies indicate that the rings are composed overwhelmingly of H$_2$O ice (see recent review by \citeA{cuzzi18b} and references therein), we find that CH$_4$ represents the largest fraction of the total MDR in our results. Previous remote observations do indicate the presence of a non water ice component to the rings, however no study has definitively determined the composition of this material. In fact, discreet searches for spectral features of CH$_4$ and NH$_3$ in ring observations found no significant amounts of either species \cite{nicholson08}. Recent in situ Grand Finale observations using Cassini's Cosmic Dust Analyzer (CDA) observed a much higher silicate fraction in the D ring than previously inferred from optical and microwave measurements \cite{hsu18}. It remains unclear why in situ measurements from INMS and CDA observe a much more significant fraction of non water ice material in the rings. Remote observations suggest that the non water material is intimately mixed within the water ice grains \cite{filacchione14, cuzzi18}. If these minor components exist in the rings as inclusions within water ice grains or as clathrates, it is possible that the resolution of existing remote observations cannot distinguish the signal from pure water ice. If water ice is deposited back onto ring particles as suspected it is also possible that the water ice is shielding the spectroscopic signature of the minor constituents within the rings, which could be part of the reason that non water ice components have evaded detection via remote observations.

H$_2$O and NH$_3$ exhibit greater variations in deposition rate than CH$_4$ from orbit to orbit, with a tenfold increase observed during orbit 291. In total, the mass deposition rate into Saturn for these three species spans from 1.7 to 8.3 $\times$ 10$^{3}$ kg/s. This is a substantial amount of incoming material and these results represent a lower limit on the amount entering Saturn from the rings since we are focusing here only on a limited set of the full mass range of the instrument. INMS detected many other species with higher mass also entering Saturn's atmosphere \cite{waite18}. Using different methods, \citeA{waite18} and \citeA{perry18} reported $>$10$^4$ kg/s of material entering the atmosphere based on INMS observations. This higher mass material is beyond the scope of this paper and will be discussed in upcoming publications. 

Recent Grand Finale results suggest that the mass of Saturn's rings is actually lower than most previous results, approximately 1.54 $\pm$ 0.49 $\times$ 10$^{19}$ kg \cite{iess19}. If we use very straightforward assumptions that the rings are able to spread over time (via viscous spreading, satellite interactions, and micrometeoritic bombardment) and continuously feed the influx of material into the atmosphere, that our influx values are constant over time, and that there are no other sources replenishing the rings, then our mass deposition results suggest that the entire ring system could be depleted in approximately 120 million years. More realistically, viscous spreading throughout the rings is not effective enough to deplete the entirety of the ring system \cite{salmon10}. The bulk of this infalling material is coming from Saturn's diffuse innermost D ring, which could result in an extremely short lifetime for the D ring and no notable effects for the more massive rings that are located further from the planet. The D ring is most likely sustained over time by material from the neighboring C ring, so it is possible that this influx is also affecting the C ring. Furthermore, observations of the D ring since the Voyager era indicate that it is rather dynamic and recent disturbances within the ring \cite{hedman07, hedman14} suggest that this large influx could be a recent and temporary development. If this is the case, the large influx of material observed in Cassini's last orbits might be a transient episode that is not indicative of typical influx values. 

\section{Conclusions}

Cassini's Grand Finale orbits performed the first ever in situ observations of Saturn's upper atmosphere. The surprisingly complex measurements returned by INMS provide us with the unique opportunity to measure the composition of the rings and understand the impact of ring rain on Saturn's equatorial atmosphere. In this study, we present further evidence of ring material inflowing into Saturn's upper atmosphere from Cassini's last few orbits, building on previous INMS results first reported in \citeA{yelle18}. The region of the INMS mass range presented here (m/z = 12 - 20 amu) indicates that CH$_4$, H$_2$O, and NH$_3$ are entering Saturn's equatorial region through dynamic ring-atmosphere interactions. We have not analyzed here the entirety of the mass range (up to 99 amu), which includes evidence for a large amount of higher mass organics also present in Saturn's upper atmosphere, and will be discussed in future work. 

Identification and quantification of these species in the returned mass spectra are made difficult by their overlapping signals in INMS measurements. This is further complicated by the fact that INMS calibration data do not exist for all species of interest and measurements from the standard NIST mass spectral library are not an identical substitute for INMS calibration values. To overcome this, we adopt a new approach to mass spectral deconvolution that uses a Monte-Carlo randomization of the peak intensity of each fragment for each species \cite{gautier19}. This method allows us to generate thousands of simulated databases to model the INMS measurements and provides a probability density for the mixing ratio of each species in the database. Retrieved mixing ratios confirm the presence of a large abundance of CH$_4$ as previously reported in \citeA{yelle18} and \citeA{waite18}, and highlight the variability of CH$_4$, H$_2$O, and NH$_3$. This variability could be connected to the volatility of these species or their differing proton affinities, which could allow for H$_2$O and NH$_3$ to more readily enter Saturn's atmosphere from the rings in a charged form at different latitudes.  

Retrieved density profiles of these species, along with Saturn's main atmospheric constituents H$_2$ and He, show that while He is in diffusive equilibrium above the homopause as expected, CH$_4$, H$_2$O, and NH$_3$ are not. Their presence in this region of the atmosphere and their nearly constant mixing ratio is consistent with an external source, which must be Saturn's rings. The measured influx for these orbits ranges from 9 to 44 $\times$ 10$^{12}$ m$^{-2}$s$^{-1}$, which translates to 1.7 to 8.3 $\times$ 10$^{3}$ kg s$^{-1}$ being deposited into Saturn's atmosphere. These results represent a lower limit on the influx from INMS observations and are thus far consistent with previous INMS measurements \cite{waite18, perry18}, and further expand on observations of ring-atmosphere interactions (e.g., \citeA{connerney84, odon17, hsu18, mitchell18}).

\acknowledgments
\change{This research was supported by Cassini Data Analysis Program Grants 80NSSC19K0903 and 16-CDAP16 2-0087.}{This research was supported by Grant 80NSSC19K0903 originally selected as part of the Cassini Data Analysis Program, and now supported by NASA's Planetary Science Division Internal Scientist Funding Program through the Fundamental Laboratory Research (FLaRe) work package, and 16-CDAP16\_2-0087.} \change{The data analyzed here is archived in the Planetary Data System and can be found at https://pds-ppi.igpp.ucla.edu/search/view/?f=yes\&id=pds://PPI/CO-S-INMS-3-L1A-U-V1.0/DATA/SATURN/2017.}{The original INMS data analyzed here is archived in the Planetary Data System and can be found at https://pds-ppi.igpp.ucla.edu/search/view/?f=yes\&id=pds://PPI/CO-S-INMS-3-L1A-U-V1.0/DATA/SATURN/2017. Data generated as a result of this analysis can be found in the Johns Hopkins University Data Archive with the following DOI https://doi.org/10.7281/T1/QGOMA0.}


%
%

\bibliography{Saturn}

%
%
%
%
%

\end{document}